\documentclass[aps,twocolumn,floatfix]{revtex4-1}

\usepackage{amsmath,amsthm,amsfonts,amssymb,times,bbm,graphicx}

\newtheorem{theorem}{Theorem}

\newtheorem{observation}[theorem]{Observation}

\newcommand{\bra}[1]{\mbox{$\langle #1 |$}}
\newcommand{\ket}[1]{\mbox{$| #1 \rangle$}}

\def\ee{\mathbbm{E}}
\def\id{\mathbbm{\mathbbm{1}}}

\def\Id{\mathbbm{1}}

\DeclareMathOperator{\diag}{diag}

\usepackage[all]{xy}

\newcommand{\xminid}[1]{
	\xymatrix@C=5mm{#1}
}
\newcommand{\Par}{\ar@{-}[r]}
\newcommand{\Pau}{\ar@{-}[u]}
\newcommand{\Pad}{\ar@{-}[d]}
\newcommand{\Pal}{\ar@{-}[l]}

\begin{document}

\title{Quantum computational webs}

\author{D.\ Gross$^{1}$ and J.\ Eisert$^{2,3}$}

\affiliation{ 
1 Institute for Theoretical Physics, Leibniz University Hannover, 30167 Hannover, Germany\\
2 Institute for Physics and Astronomy,
University of Potsdam, 14476 Potsdam, Germany\\
3 Institute for Advanced Study Berlin, 14193 Berlin, Germany}

\date{\today}

\begin{abstract} 
We introduce the notion of quantum computational webs: These are
quantum states universal for measurement-based computation 
which can
be built up from a collection of simple primitives. The primitive
elements---reminiscent of building blocks in a  construction 
kit---are (i) one-dimensional states (``computational quantum wires'')
with the power to process one logical qubit and (ii) suitable
couplings which connect the wires to a computationally universal
``web''. All elements are preparable by nearest-neighbor interactions
in a single pass, as are accessible in a number of physical architectures. 
We provide a complete classification of 
qubit wires---this being the first instance where a physically 
well-motivated class
of universal resources can be fully understood.
Finally, we sketch possible realizations in superlattices, and explore
the power of coupling mechanisms based on Ising or
exchange-interactions.
\end{abstract}

\maketitle

It is an intriguing fact that one can perform universal quantum
computation just by performing local measurements on certain quantum
many-body systems 
\cite{Oneway,Cluster,Survey,GS,Tele,OneWayVB,MBC,MBC2}.
Despite enormous interest in this 
phenomenon, our
understanding of which quantum systems offer a quantum computational
speed-up and which do not is still rudimentary. Indeed, for years the
only states known to be universal for quantum computation by
measurements were the cluster state and very close 
relatives \cite{Oneway,Cluster,Greiner}.
This was unsatisfactory both from a fundamental point of view and for
experimentalists aiming to tailor resource states to their physical
systems in the lab.
In Refs.\ \cite{MBC,MBC2} a framework for the construction of new
schemes for MBQC was introduced (further applied e.g. in Refs.\
\cite{SL,M}). There, it was shown that many of the singular properties
of the cluster are \emph{not} necessary for a computational 
speed-up---hence weakening the requirements for MBQC. This newly found
flexibility notwithstanding, it has been established that universality
is a rare property among quantum many-body states~\cite{useless}.
Therefore, it would be very desirable to obtain a full
classification of the relatively few states which are universal. While
the unqualified problem still seems daunting, we show in this work
that under reasonable, physically motivated constraints, a complete
understanding is possible.

\begin{figure}
\includegraphics[width=7.2cm]{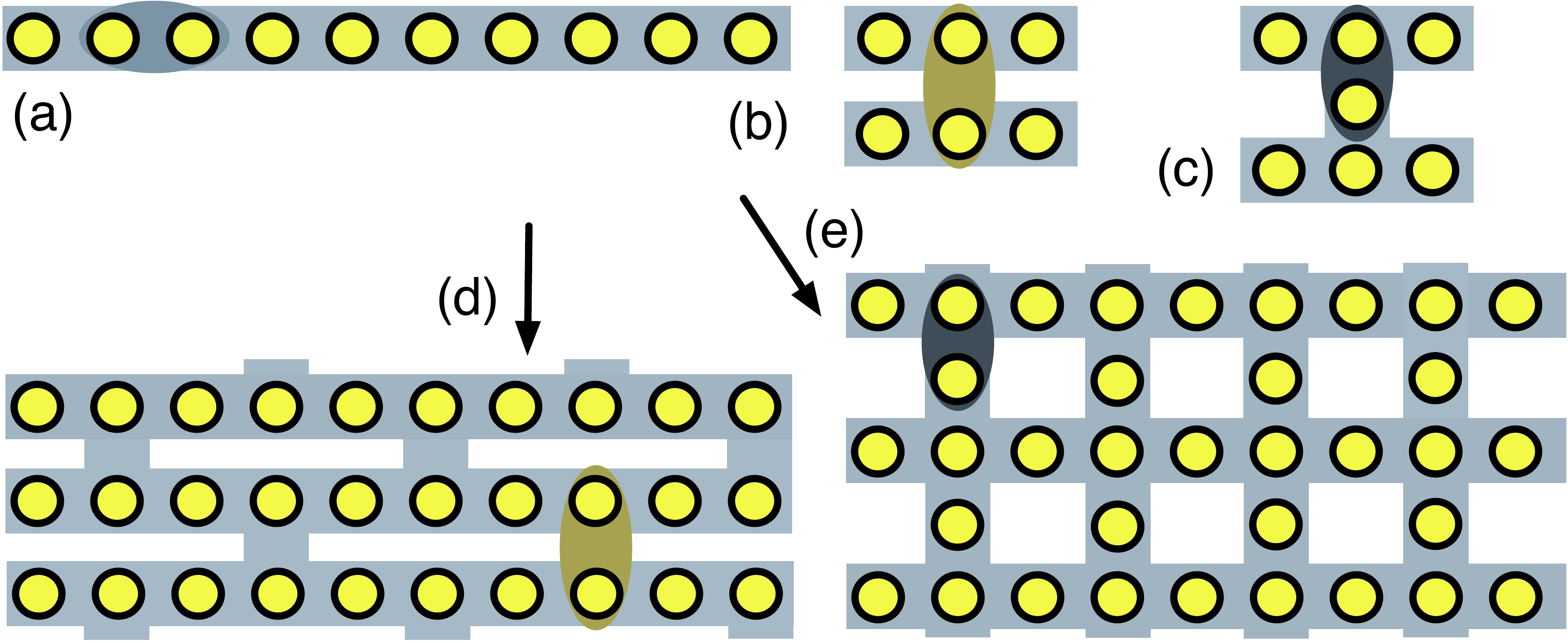}
\caption{\label{sketchFig}
Sketch of the primitives from which one can built up new models for
computing: (a) A general quantum computational wire. Two different
coupling schemes based on (c,e) an Ising-type interaction or (b,d)
Heisenberg-type or exchange interaction (the latter being defined for
cluster wires).}
\end{figure}

The basic idea is to break up resource states into smaller primitives,
which are more amenable to analysis. Indeed, most known states
universal for MBQC come in two versions: (i) states on a 1-D chain of
qubits, 
which have the ability to transport and process one logical
qubit worth of quantum information
\cite{Oneway,MBC,MBC2,SL,M},  
and (ii) 2-D versions, obtained by
suitably entangling several 1-D strands. We will refer to such 1-D
states as \emph{quantum computational wires}.
They form the
measurement-based equivalent of a single qubit. Likewise, the
\emph{couplings} used to form truly universal 2-D resources 
(referred to as {\it quantum computational webs})
are the 
analogues of entangling unitaries in the gate model. Splitting the
analysis of universal states into wires and couplings has two
advantages: (i) the primitives are far easier to understand than the
compound state they give rise to and (ii) in a manner reminiscent of a
construction kit, wires and couplings may be freely combined to form
diverse sets of universal resources (c.f. Fig.~\ref{sketchFig}).

{\it Full classification of qubit wires.} For most of what follows
we focus on qubit systems, for which we can provide a full theory. 
We impose the physically reasonable requirement that wires can be
build up from product states by means of nearest-neighbor
interactions $U=e^{-itH^{(i,i+1)}}$ in a single translationally
invariant pass. The physical realizations we have in mind here are
atoms in an optical lattice as in an ``atomic sorting device'' \cite{Meschede},
settings exploiting optical superlattices \cite{SL,Superlattice}, or 
other architectures such as ones involving interacting
quantum dots \cite{Vandersypen} or instances of
networks \cite{Kimble}. More specifically, by 
a {\it qubit computational wire} we mean
\begin{itemize}
\item [(i)] a family of pure states $|\phi_n\rangle$ of a $n$-qubit spin
chain,
\item[(ii)] preparable from a product state 
$|0,\dots, 0\rangle$
by the sequential action of a unitary gate $U$
\begin{equation}\label{prep}
	|\psi_n\rangle = U^{(n,n-1)}\dots
	U^{(3,2)}
	U^{(2,1)}|0,\dots, 0\rangle.
\end{equation}
\item[(iii)] In the limit of large $n$, the 
entanglement
between the left and the right half of the chain 
(in the sense of an ``area law'') 
approaches one ebit.
\end{itemize}

These axioms may seem surprisingly weak: earlier, we loosely
characterized computational wires as states with the power to
``transport and process one logical qubit''.  It is one central result
of this work that any state fulfilling (i)--(iii) is automatically
useful for information processing. Below, we will explain
and prove the following complete classification of qubit wires up to
local basis changes:

\begin{observation}[Classification of qubit wires]\label{qubit}
There is a three-parameter family of 
computational qubit wires.
A wire is specified by an 
\begin{itemize}
\item[(a)]``always-on operation''
$W\in SU(2)$, acting on correlation space (see below) after
every step, independent of the basis chosen or
the measurement outcome, and a 
\item[(b)]``by-product angle''
$\phi$, specifying how sensitive the 
resource is to the inherent 
randomness of measurements.
\end{itemize}
\end{observation}

\begin{figure}
\includegraphics[width=7.9cm]{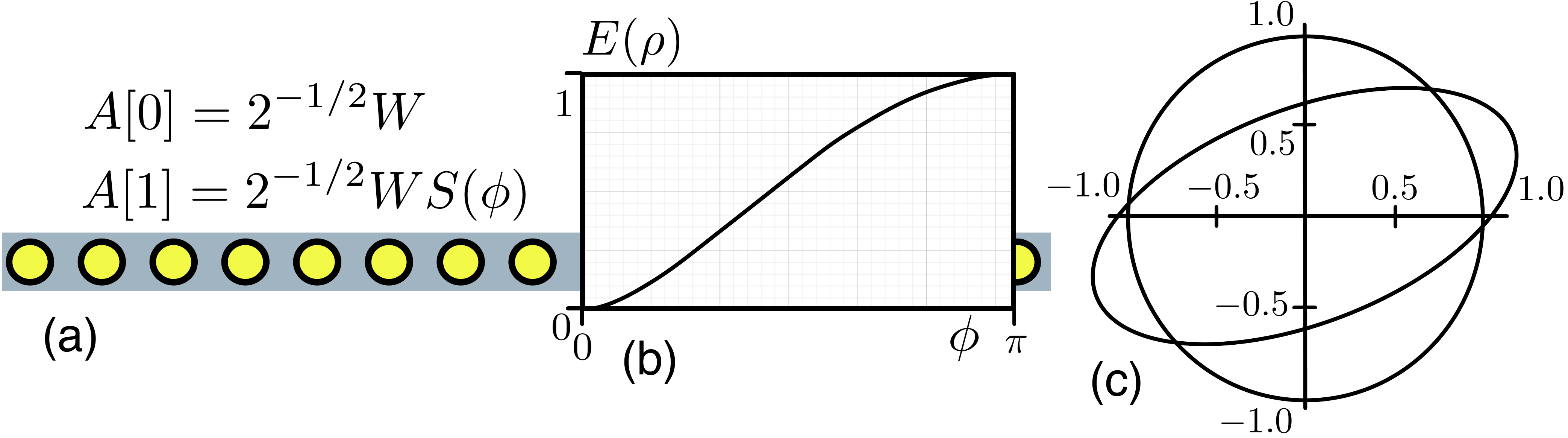}
\caption{\label{wireFig}
(a) Normal form of a qubit wire, (b) entropy of 
entanglement of a single site as a function of the by-product
angle, and (c) trajectory of all operations realizable in a wire with
$\phi=\pi$ (circle) and $\phi=\pi/2$ (ellipse). Every point $\sqrt p
e^{i\delta}$ on the curve corresponds to the operation $S(-2\delta)$,
realizable with probability $p$.
}
\end{figure}

To make sense of this statement, first note that any $\ket{\psi_n}$
has a \emph{matrix product state} (MPS) representation
\cite{MPS,MBC,MBC2}: 
\begin{eqnarray}\label{general}
	|\psi_n\rangle=
	\sum_{x_1,\dots ,x_n}
	\bra{x_n} A[x_{n-1}] \dots A[x_1]|0\rangle\,
	|x_1,\dots, x_n\rangle,
\end{eqnarray}
where 
$x_i\in\{0,1\}$, and $A[0],A[1]$ are $2\times 2$-matrices. (Eq.\
(\ref{general}) follows from Eq.~(\ref{prep}) by setting $A[x]_{i,j}=
\langle i,x |U |0,j\rangle$.) The auxiliary two-dimensional vector
space the matrices $A[0/1]$ act on is called \emph{correlation space}.
We recall very briefly the basic idea of Refs.~\cite{MBC,MBC2}.
Let $\ket{\phi^{(i)}} = c^{(i)}_0\ket0+c^{(i)}_1\ket1$ be a local state
vector and set $A[\phi^{(i)}]=\bar{c}^{(i)}_0 A[0]+\bar{c}^{(i)}_1A[1]$.
Then
\begin{equation*}
	(\bra{\phi^{(1)}}\otimes\dots\otimes\bra{\phi^{(n)}})	
	\ket{\psi_n}
	=\bra{\phi_n} A[\phi_{n-1}] \dots A[\phi_1]\ket 0.
\end{equation*}
Hence, a local measurement with outcome corresponding to $\ket{\phi_i}$
is connected with the action of the operator $A[\phi_i]$ on
correlation space. MBQC can be understood completely in terms of this
relation between local measurements and logical computations on
correlation space \cite{MBC,MBC2}. With these notions, the precise
statement of Obs.~\ref{qubit} is that any wire allows for an
MPS representation with matrices
\begin{equation}\label{eqn:mpsRep}
	B[0]=2^{-1/2} W, \qquad B[1]=2^{-1/2} W S(\phi),
\end{equation}
where $S(\phi)=\diag(e^{-i\phi/2},e^{i\phi/2})$, see Fig.\ 
\ref{wireFig}(a). (That is to say, any matrix arising from 
Eq.\ (\ref{general}) can be brought into this form by a suitable 
rescaling and conjugation, see below).

Obs.~\ref{qubit} goes a long way towards understanding the structure
of qubit wires. Assume that we measure site by site in the
computational basis. By Eq.~(\ref{eqn:mpsRep}), 
at every step the same
{\it ``always-on''-operation} $W$ will be applied to the correlation
space, irrespective of the measurement outcome.  Some tribute
must be paid to the random nature of quantum measurements. It comes in
the form of the \emph{by-product} operation $S(\phi)$, acting on the
correlation system in case the ``wrong'' measurement outcome (``\ket1'',
instead of ``\ket0'') is obtained. It is remarkable that this penalty is
described by a single parameter: the \emph{by-product angle} $\phi$ \cite{Note}.

{\it Examples of qubit wires.--} The paradigmatic qubit wire is the
{\it cluster state}. Here, $W=H$, the Hadamard gate,
and $\phi=\pi$, the highest possible value \cite{Hadamard}.
We can thus put two well-known properties of the cluster into a more
general context: (i) in every step a Hadamard gate $H$ is applied to
the logical qubit and (ii) a ``wrong'' measurement outcome causes the
application of an extra $S(\pi)\simeq \sigma_z$ gate on correlation
space.

Another interesting new resource where the role of the by-product
angle can be highlighted is the $T${\it -resource}, named after the
common notation $T=S(\pi/2)$ for a phase gate. Here, we take $W=H$ (as
for the cluster), but the by-product angle is just $\phi=\pi/2$ (so
that a measurement in the computational basis gives rise to either $H$
or $HT$).  This qubit wire has non-maximal entropy of
entanglement of a single site w.r.t.\ the rest of the lattice. The
intuitive explanation is that $T$ is ``close'' to the identity, so the
state of the correlation system (and hence the rest of the chain) does
not strongly depend on the outcomes of local measurements on any given
site.

The proof of Obs~\ref{qubit} will make repeated use of the theory of
MPS's \cite{MPS} and of qubit channels \cite{Ruskai}. Any MPS can be
represented with matrices s.t.\ $A[0]^\dagger A[0]+A[1]^\dagger
A[1]=\id$ \cite{MPS}.  The matrices give rise to a trace-preserving
channel $\rho\mapsto \ee(\rho)= \sum_x A[x] \rho A[x]^\dagger$.
Assuming that $\ee$ has a spectral gap \cite{gap} the half-chains
share one ebit of entanglement iff $\ee$ is unital \cite{MPS}. In this
case, it follows easily from Ref.\ 
\cite{Ruskai} that $\ee(\rho)=p_0 U_0
\rho U_0^\dagger + p_1 U_1 \rho U_1^\dagger$, with suitable $U_i\in
SU(2)$. From the basic theory of quantum channels, we know that there
is a unitary $V\in SU(2)$ such that $p_i^{1/2} U_i=\sum_j V_{i,j} A[j]$.
That being nothing but the transformation rule for MPS representations
under local basis change, we conclude that there is a basis in which
$\ket{\psi_n}$ is represented with matrices $A'[i]=p_i^{1/2} U_i$. Next, an
MPS does not change if both matrices are conjugated by the same
operator $X$. There is an $X\in SU(2)$ such that $X U_0^\dagger U_1
X^\dagger = e^{i\alpha} S(\phi)$ for $\alpha,\phi\in\mathbbm{R}$.
Setting $W=X U_0 X^\dagger$ and $B[i]=X A'[i] X^\dagger$ implies
$B[0]=p_0^{1/2} W, B[1]=p_1^{1/2}  e^{i\alpha} W S(\phi)$. Performing the local
basis change $\ket 1 \mapsto e^{i\alpha}\ket 1$ if necessary, we may
assume that $\alpha=0$. The fact that $p_0, p_1$ can be chosen to be
$1/2$ will be explained below in a more general context. 
Conversely, any MPS with matrices as in Eq.~(\ref{eqn:mpsRep}) is a
qubit wire. A translationally invariant preparation scheme can
easily be derived by inverting the construction below
Eq.~(\ref{general}).

{\it Computation with qubit wires.} So far we have shown that one
can implement \emph{some} unitary operation in a quantum wire, 
i.e.\ 
\emph{transport} quantum information. In order to \emph{process} it,
one must have some freedom to choose which operation to apply.
It will turn out---rather surprisingly---that two coincidences conspire
to make any qubit quantum wire useful for that purpose.
To that end, consider the one-parameter family of bases
\begin{equation*}
	\ket{0_\theta}=\sin(\theta)\ket0+\cos(\theta)\ket1,\>
	\ket{1_\theta}=\cos(\theta)\ket0-\sin(\theta)\ket1.
\end{equation*}
One may check directly that the operations $A[0_\theta]\propto
W(\sin\theta\Id + \cos\theta S(\phi))$ are unitary up to scaling.  The
two unexpected coincidences are: (i) for any quantum wire, there is a
continuous family of projections which give rise to unitary evolution
and (ii) the set these projections includes entire bases---so that
measuring in these bases corresponds to a unitary logical computation
regardless of the outcome.
\begin{observation}[Unitary evolution]
	For any computational wire, 
	a measurement in any basis from the
	one-parameter set 
	$\{\ket{0_\theta},\ket{1_\theta}\}$
	induces a unitary evolution in correlation space.
\end{observation}
Let us investigate the realizable unitaries. Clearly, $A[0_\theta]$
has the form $W U(\theta,\phi)$, where $U(\theta,\phi)$ is  a diagonal
matrix with eigenvalues $\lambda_{\pm}=\sin(\theta)+\cos(\theta)
e^{\pm i\phi/2}$. Let $\delta= \arg(\lambda_+)$ and $p
=|\lambda_+|^2$. Then $U(\theta,\phi)=\sqrt p\,S(-2\delta)$ and basic
MPS theory yields that the corresponding measurement outcome is
obtained with probability $p$. For fixed $\phi$, the set of phase
gates $S(-2\delta)$ thus realizable forms an ellipse,
see Fig.\ 
\ref{wireFig}(c)  \cite{Note2},
in the complex
plane with parametrization
\begin{eqnarray*}
	(\text{re} \lambda_+(\theta,\phi), \text{im}
	\lambda_+(\theta,\phi))^T 
	&=&
	\left(
	\begin{array}{cc}
		1&\cos\phi/2 \\
		0&\sin\phi/2 
	\end{array}
	\right)
	\left(
	\begin{array}{cc}
		\sin\theta\\
		\cos\theta
	\end{array}
	\right).
\end{eqnarray*}
\begin{observation}[Phase gates]
	In any computational wire, an \emph{arbitrary phase gate}
	$S(\delta)$ can be implemented in a single step. 
\end{observation}
Leaving aside the issue of randomness for a moment, we see that one
can realize any unitary of the form $U=W S(\delta_n)W
S(\delta_{n-1})\dots W S(\delta_1)$ for some $n$. Invoking assumption
\cite{gap}, every $U\in SU(2)$ is of that form.
\begin{observation}[Universal rotations]\label{obs:rotations}
Except from a set of measure zero, all
computational qubit wires allow for the
implementation of any unitary $U\in SU(2)$ 
in correlation space.
\end{observation}

{\it Local properties.}  From MPS theory \cite{MPS}
one finds that the reduced state of a single site 
far away from the boundary is given by
$\rho=\sum_{i,j} \text{tr}(A[i]^\dagger A[j]) \,\ket i \bra j /2$.
Explicitly:
\begin{equation}\label{eqn:densityPhi}
	\rho= 
	\left(
		\begin{array}{cc}
			1 & \cos\phi/2 \\
			\cos\phi/2 & 1
		\end{array}
	\right)/2.
\end{equation}
Interestingly, we see that the 
always-on operation $W$ does not affect the
local properties of the state. One can hence conclude
(see Fig.\ \ref{wireFig}(b)):

\begin{observation}[Small entanglement in wires]
	Computational wires with arbitrarily low 
	local entanglement exist.
\end{observation}

{\it Compensating randomness.} In the above classification, we required
from a ``qubit wire'' to allow for ``transporting and processing one logical qubit''.
We yet also need to clarify how to deal with the inherent randomness of quantum measurements.  
If the always-on term $W$ and the by-product operator $S(\phi)$ generate a finite
group $\mathcal{B}$, there is a simple and efficient possibility to cope with
randomness, introduced in Ref.\ \cite{MBC}: Suppose we would like to
implement $WS(\delta)$, but instead obtain a measurement outcome which
causes $WS(\delta')$ to be realized.  Now, by measuring several
consecutive sites in the computational basis, we effectively implement
a random walk on the finite group $\mathcal{B}$ in correlation space.
This random walk will visit any element of $\mathcal{B}$ after a
finite expected number of steps. We will hence obtain
$W^{-1}\in\mathcal{B}$ after several steps, yielding a total evolution of
$W^{-1} W S(\delta')=S(\delta')$. Then, one tries to implement
$S(-\delta'+\delta)$, which is possible by Obs.~\ref{obs:rotations}
\cite{NewNote}.
It remains to be shown how logical information in the correlation system
can be prepared and read out. As for preparation, note that
$A[2^{-1/2}(\ket0 - e^{\phi/2}\ket1)]\propto \ket 1 \bra 1$ has rank one.
Hence, if after a local measurement the outcome corresponding to
$2^{-1/2}(\ket0 - e^{\phi/2}\ket1)$ is obtained, the correlation system
will be in $\ket 1$, irrespective of its previous state---so 
preparation is possible. A read-out scheme can be devised
along these lines.

\begin{observation}[Preparation and readout]
For any qubit wire, one can efficiently prepare the correlation system
in a known state and read out the latter by local measurements.
\end{observation}

\begin{figure}
\includegraphics[width=9.1cm]{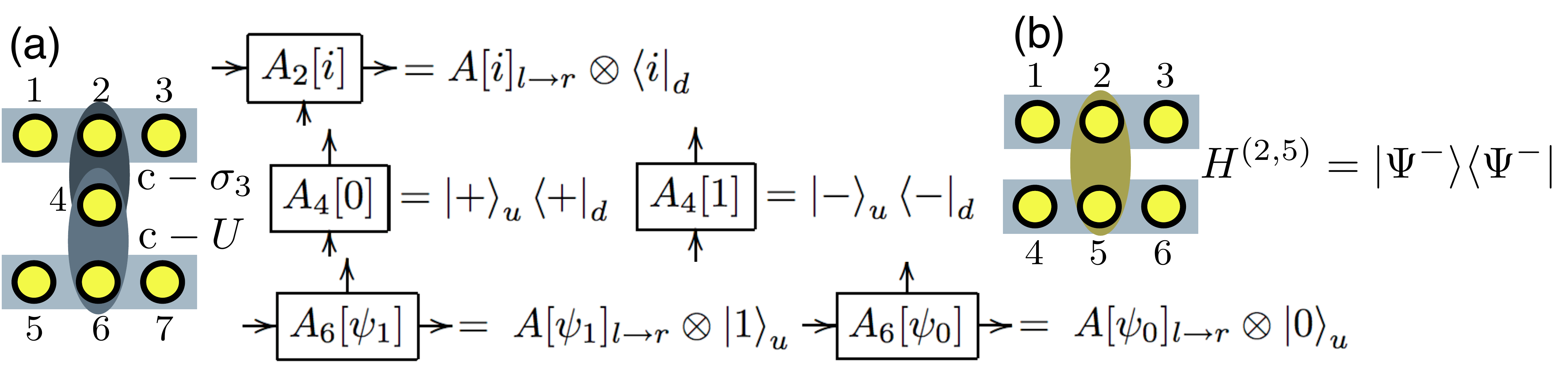}
\caption{\label{IsingFig}
(a) Universal coupling scheme based on two 
(Ising-type)
controlled-unitary
gates for arbitrary qubit
wires. For completeness, we also state
the {\it tensor network} in the language of 
Refs.\ \cite{MBC,MBC2}. (b) A coupling of cluster wires based on an
exchange interaction.}
\end{figure}

{\it Ising coupling.} All wires introduced so far can be coupled to
form a 2-D state, universal for quantum computation. Remarkably, there are
several coupling schemes, which work equally well for \emph{all} 1-D
states so far introduced. Space limitations require us to describe
only one and be somewhat sketchy (however, all central points are
explained; see Ref.\ \cite{upcoming} for further details). 
The coupling scheme, depicted in Fig.\ \ref{IsingFig}a,
is based on a setting where $\{1,2,3\}$ and
$\{5,6,7\}$ belong to any wire and $4$ has been prepared in
$2^{-1/2}(|0\rangle+|1\rangle)$. One now entangles sites $\{2,4\}$ and
sites $\{4,6\}$ via Ising interactions in a suitable bases. More
concretely, one performs a controlled-$\sigma_z$ gate ($CZ^{(2,4)}$)
between site $2$ and site $4$ and then applies
\begin{equation*}
	W^{(6)} CZ^{(4,6)} (W^{(6)})^\dagger,\>
	W=\left(
	\begin{array}{cc}
	1 & 1 \\
	e^{i\phi/2} & -e^{-i\phi/2}
	\end{array}
	\right)2^{-1/2}
\end{equation*}
between systems $4$ and $6$. To decouple the wires, just measure $4$
in the computational basis. In case of the $\ket0$-outcome, we have
un-done the coupling; a $\ket1$-outcome brings us back to the original
state, up to the action of $\sigma_z$ on site $2$ and $W\sigma_z
W^\dagger$ on $6$.  To perform an entangling gate, one measures $6$ in
the $\sigma_z$-basis and $4$ in the $\sigma_x$-basis, getting outcomes
$x_4, z_6\in\{0,1\}$ respectively. Let us assume that $x_4+z_6$ is even.
Choose $\gamma,\varepsilon$ such that
$e^{i\varepsilon/2}\sin\gamma=1/2(1-e^{i\phi})$, 
and let $\delta$ be the solution to $|\cos\delta| =
|\sin\delta\sin\gamma+ \cos\delta\cos\gamma e^{i\phi/2}|$
(which always exists). Finally,
measure site $2$ in the basis
$|\psi_0\rangle=e^{-i\varepsilon}\sin\delta|0\rangle+
\cos\delta |1\rangle$,
$|\psi_1\rangle=
- e^{-i\varepsilon}\cos\delta|0\rangle+
\sin\delta|1\rangle$.
A lengthy---but by these definitions fully specified---calculation shows
that if we get the $|\psi_0\rangle$ outcome, then one implements 
the unitary entangling operation
\begin{eqnarray}
	V&=&W|0\rangle \langle 0|\otimes (\cos(\delta)A[1])\\
	&+&
	W|1\rangle\langle 1|\otimes (\sin(\delta)\sin(\gamma)
	A[0] + \cos(\delta) \cos(\gamma)A[1])\nonumber
\end{eqnarray}
between the upper and lower correlation spaces. The orthogonal
outcome and the case of odd $x_4+z_6$ may be treated
analogously.

\begin{observation}[Ising-type coupling] Arbitrary
qubit wires can be coupled with suitable phase gates.
\end{observation}

We use the remaining paragraphs to give an outlook on further results
and ideas.

{\it Exchange interaction coupling.} Using the ideas presented
above, one may check that cluster wires can be coupled together using
an exchange interaction Hamiltonian:
$H_{ex}=\ket{\Psi^-}\bra{\Psi^-}$, where
$\ket{\Psi^-}=2^{-1/2}(\ket{0,1}-\ket{1,0})$.
The topology used here is a ``hexagonal lattice with additional
spacings'', see Figs.\ \ref{sketchFig}(b). The coupling operation used
to obtain a universal resource is given by $U=e^{i \pi /2 H_{ex}}$ \cite{upcoming}.
 
 \begin{observation}[Exchange interaction coupling]
An exchange interaction Hamiltonian can be used to couple cluster
wires.
\end{observation}
{\it Bose-Hubbard-type and continuous-variable wires.} Widening our
scope beyond qubits, we look at bosons in optical superlattices
\cite{SL,Superlattice}, subject to an {\it Bose-Hubbard} interaction (compare also Ref.\ \cite{Loss}).
Consider the situation where the potential forms a string of
double-wells, with the right site of each double-well occupied by a
single particle, $\ket{\Psi(t=0)}= |0,1,\dots, 0,1\rangle$.  In a
first step, one lets the two sites of each double-well interact with
$H= a_L^\dagger a_R + a_R^\dagger a_L$ for time $t=\pi/4$, leading to
pairs in the state $|0,1\rangle+ i|1,0\rangle$. Secondly---in the
fashion of a quantum {\it cellular automaton}---one shifts the
superlattice, so that neighboring pairs which have not previously
interacted are subjected to the Hamiltonian above. One obtains a
globally entangled state with up to three excitations per site and
entropy of entanglement between half-chains of up to $E(\rho)= 1.725$.
Assuming the power to perform 
tilted measurements in the particle number basis (or making
use of suitable internal degrees of freedom) 
it is easily checked that this ``Bose-Hubbard wire'' allows for the
transport of one logical qubit, and arbitrary rotations along one
axis.  This is an example of a primitive where the local Hilbert space
dimension is in principle infinite. Further steps towards
continuous-variable (CV) schemes could be done by considering
correlation spaces where only a subspace of superpositions of finitely
many coherent states is occupied, such that the correlation space is
still finite-dimensional. The framework established here forms a
starting point to study such CV computational schemes.
\begin{observation}[Bose-Hubbard wires]
Suitable states preparable by Bose-Hubbard interactions in
superlattices allow for 
transport of one logical qubit.
\end{observation}

{\it Summary.} We have introduced a toolbox of primitives for
constructing new quantum computational schemes. For the qubit case, we
provide a full classification. The results constitute a further
step towards the goal of understanding what is ultimately needed for
quantum computation and what degree of freedom there is in designing
computational schemes.

{\it Acknowledgements.} We thank the EU (QAP, QESSENCE, COMPAS, CORNER)
and the EURYI scheme for support and A.\ Kay, I.\ Bloch and
L.\ Vandersypen for discussions.


\begin{thebibliography}{99}

\bibitem{Oneway}
        R.\ Raussendorf and H.-J.\ Briegel,
        Phys.\ Rev.\ Lett.\ {\bf 86},  5188 (2001);
	R.\ Raussendorf and H.-J.\ Briegel, 
	Quant.\ Inf.\ Comp.\ {\bf 6}, 433 (2002).

\bibitem{Cluster}
       H.-J.\ Briegel and R.\ Raussendorf,
        Phys.\ Rev.\ Lett.\ {\bf 86}, 910 (2001).
 
\bibitem{Survey}        
        M.\ Hein, W.\ D{\"u}r, J.\ Eisert,
        R.\ Raussendorf, M.\ Van den Nest, and
        H.-J.\ Briegel,
        quant-ph/0602096.

\bibitem{GS}
        M.\ Hein, J.\ Eisert, and H.-J.\ Briegel,
        Phys.\ Rev.\ A {\bf 69}, 062311 (2004);
        R.\ Raussendorf, D.\ E.\ Browne, and H.-J.\ Briegel,
        ibid.\ {\bf 68}, 022312 (2003);
        D.\ Schlingemann and R.\ F.\ Werner,
        ibid.\ {\bf 65}, 012308 (2002);
        P.\ Walther et al., Nature {\bf 434}, 169  (2005).
     
\bibitem{Tele}      
     	P.\ Aliferis and D.\ W.\ Leung, Phys.\ Rev.\ A {\bf 70}, 062314
	(2004); P.\ Jorrand and S.\ Perdrix, quant-ph/0404125 (2004).
	 
\bibitem{OneWayVB}		
	F.\ Verstraete and J.\ I.\ Cirac,
	Phys.\ Rev.\ A {\bf 70}, 060302(R) (2004).		
           		
\bibitem{MBC}
	D.\ Gross and J.\ Eisert, Phys.\ Rev.\ Lett.\ 
	{\bf  98}, 220503 
	(2007).

\bibitem{MBC2}	
	D.\ Gross, J.\ Eisert, N.\ Schuch, and D.\ Perez-Garcia,
	Phys.\ Rev.\ A {\bf 76}, 052315 (2008).

\bibitem{Greiner}
	O.\ Mandel et al., Nature {\bf 425}, 937 (2003).
	
\bibitem{useless}
	D.\ Gross, S.\ T.\ Flammia, and J.\ Eisert, Phys.\ Rev.\ Lett.\ {\bf 102}, 190501 (2009);
	M.\ J.\ Bremner, C.\ Mora, and A.\ Winter, ibid.\ {\bf 102}, 190502 (2009).
			
\bibitem{SL}
	B.\ Vaucher, A.\ Nunnenkamp, and D.\ Jaksch,
	New J.\ Phys.\ {\bf 10}, 023005 (2008). 

\bibitem{M}
	G.\ K.\ Brennen and A.\ Miyake, 
	Phys.\ Rev.\ Lett.\ {\bf 101}, 010502 (2008). 	

\bibitem{Meschede}
	Y.\ Miroshnychenko et al.,
	Nature {\bf 442}, 151 (2006).
	
\bibitem{Superlattice}
	P.\  Barmettler at al., Phys.\ Rev.\ A {\bf 78}, 012330 (2008);
	S.\ Peil et al., ibid.\  {\bf 67}, 051603 (2003);
	 I.\ Bloch, Nature {\bf 453}, 1016 (2008).
	 
\bibitem{Loss}
	M.\ Borhai and D.\ Loss, Phys.\ Rev.\ A {\bf 71}, 034308 (2005).

\bibitem{Vandersypen}
	R.\ Hanson et al.,
	Rev.\ Mod.\ Phys.\ {\bf 79}, 1217 (2007).
	 	
\bibitem{Kimble}
	H.\ J.\ Kimble, 
	Nature {\bf 453}, 1023 (2008).
			
\bibitem{MPS}
	M.\ Fannes, B.\ Nachtergaele, and R.\ F.\ Werner,
	Commun.\ Math.\ Phys.\ {\bf 144}, 443 (1992);
	D.\ Perez-Garcia, F.\ Verstraete, M.\ M.\ Wolf, and
	J.\ I.\ Cirac, Quant.\ Inf.\ Comp.\ 
	{\bf 7}, 401 (2007).

\bibitem{Ruskai}
	M.\ B.\ Ruskai, S.\ Szarek, and E.\ Werner,
	Lin.\ Alg.\ Appl.\ {\bf 347}, 159 (2002).
	
\bibitem{Note}
	Note that for the definition of a qubit wire as such, we do not require being able to
	compensate randomness of outcomes by exploiting a finite group
	structure of the by-product operators.
	
\bibitem{Hadamard}	
	Note that our
	definition differs from the conventional one by the action of a local
	Hadamard gate on every site.
	
\bibitem{gap}
	Away from (and independently of) the boundaries, an MPS is 
	completely specified by the
	matrices Eq.~(\ref{eqn:mpsRep}) iff the map $\ee$ 
	has a spectral gap \cite{MPS}. 
	This is true iff $W$ is neither diagonal nor equal to $\sigma_x$. 
	We will always implicitly assume this generic situation. 

\bibitem{Note2}	
	We can now prove the earlier claim that in the normal form
	Eq.~(\ref{eqn:mpsRep}) the weights of the two matrices may be chosen
	to be equal. That follows from the fact that there are two
	perpendicular vectors intersecting the ellipse at the same length
	$\sqrt p$. 
	
\bibitem{NewNote}	 
	 More generally, the method sketched above may be implemented as soon
	as there is some basis $\{\ket{0_\theta},\ket{1_\theta}\}$ s.t.\
	$A[0_\theta], A[1_\theta]$ generate a finite group (up to scalars). It
	can be shown that whenever one such basis exists, there is a one-parameter
	set of bases with the same property \cite{upcoming}. This gives rise
	to continuous families of wires in which randomness can be
	compensated by the same method.
	
\bibitem{upcoming}
	D.\ Gross and J.\ Eisert, 
	in preparation.
	
\end{thebibliography}
\end{document}